\documentclass{article}

\usepackage{arxiv}
\usepackage{graphicx}
\usepackage[justification=centering]{caption}
\usepackage{multirow}
\usepackage{indentfirst}
\usepackage{amsmath}
\usepackage{algorithm}
\usepackage{algorithmic}
\usepackage{setspace}
\usepackage[utf8]{inputenc} 
\usepackage[T1]{fontenc}    
\usepackage{hyperref}       
\usepackage{url}            
\usepackage{booktabs}       
\usepackage{amsfonts}       
\usepackage{nicefrac}       
\usepackage{microtype}      
\usepackage{lipsum}
\title{NegDL: Privacy-Preserving Deep Learning Based on Negative Database}

\author{
 Dongdong Zhao, Pingchuan Zhang, Jianwen Xiang\thanks{Corresponding author: Jianwen Xiang (jwxiang@whut.edu.cn).}, Jing Tian\\
 Hubei Key Laboratory of Transportation\\
Internet of Things\\
School of Computer and Artificial Intelligence\\
Wuhan University of Technology\\
Wuhan, Hubei, China\\
}
\begin{document}
\maketitle
\begin{abstract}
{In the era of big data, deep learning has become an increasingly popular topic. It has outstanding achievements in the fields of image recognition, object detection, and natural language processing et al. The first priority of deep learning is exploiting valuable information from a large amount of data, which will inevitably induce privacy issues that are worthy of attention. Presently, several privacy-preserving deep learning methods have been proposed, but most of them suffer from a non-negligible degradation of either efficiency or accuracy. Negative database (\textit{NDB}) is a type of data representation which can protect data privacy by storing and utilizing the complementary form of original data. In this paper, we propose a privacy-preserving deep learning method named NegDL based on \textit{NDB}. Specifically, private data are first converted to \textit{NDB} as the input of deep learning models by a generation algorithm called \textit{QK}-hidden algorithm, and then the sketches of \textit{NDB} are extracted for training and inference. We demonstrate that the computational complexity of NegDL is the same as the original deep learning model without privacy protection. Experimental results on MNIST, and CIFAR-10 benchmark datasets demonstrate that the accuracy of NegDL is comparable to the original deep learning model in most cases, and it is also comparable to the method based on differential privacy.}
\end{abstract}


\section{Introduction}
In the past decade, deep learning has achieved great attention from both academic community and industry due to its capability on exploiting useful knowledge from large-scale data. Deep learning has been widely used in variety of fields and has made many remarkable breakthroughs. However, few studies have focused on the importance of the privacy issue that becomes more important than ever since a large amount of data are involved. For example, medical data may contain the private data from patients, e.g., their diseases, family history and DNA sequence. If these data are not well protected during deep learning in real-world scenarios, the sensitive information may be leaked and personal privacy will be threatened.

In recent years, several methods for privacy-preserving deep learning have been proposed. however, most of them suffer from a non-negligible degradation of either efficiency or accuracy. Specifically, the methods based on differential privacy [1] protect data privacy by adding noise, which will influence the data precision and utility. The methods based on homomorphic encryption [2] usually require high computational cost and the efficiency would become unbearable in the scenarios with large-scale data. 

Negative database (\textit{NDB}) is a form of information representation, which is inspired by the Negative Selection Mechanism in Artificial Immune System [3]. \textit{NDB} stores the information in the complementary set of database (\textit{DB}) to achieve privacy protection, which can also support operations like insert, delete, update, and select as traditional databases. It has been proven to be an \textit{NP}-hard problem to reverse the negative database to recover original data. Moreover, it supports a rough distance estimation. These characteristics make it applicable to many fields for protecting privacy. However, the application of \textit{NDB} to privacy-preserving deep learning has not been addressed by the literatures so far. Therefore, in this paper, we propose an approach to use \textit{NDB} to protect data privacy during deep learning, named Negative Deep Learning (NegDL). In NegDL, the data are firstly converted into \textit{NDB}s by using the \textit{QK}-hidden algorithm [4]; to  compress the size of NDBs but retain necessary information for the neural network model, the sketches of \textit{NDB}s are extracted and inputted into the neural network; the results of activation functions are then estimated from sketches; the remaining steps are the same to general deep learning models.

The main contributions of this paper can be summarized below:

Our contributions are as follows:

\begin{itemize}
\item A method called NegDL for privacy preserving deep learning based on negative databases has been proposed.
\end{itemize}

\begin{itemize}
\item The sketches instead of negative databases are used to enhance the efficiency, and a model for calculating activation functions based on sketches has been proposed.
\end{itemize}

\begin{itemize}
\item Several experiments have been conducted on two datasets: MNIST [5], CIFAR-10 [6], and the accuracy of the proposed method reaches 99.04\%, 95.56\%, respectively, which demonstrates that the proposed method could effectively protect data privacy while maintaining most of the performance of deep learning models.. 
\end{itemize}

The rest of this paper is organized as follows: Section 2 discusses related work; Section 3 presents the background of \textit{QK}-hidden algorithm; the proposed NegDL is presented in 4; Section 5 shows the experimental results on both the accuracy and security of NegDL; and Section 6 concludes this work.

\section{Related work}

\subsection{Privacy-Preserving deep learning}

At present, several privacy-preserving deep learning methods have been proposed. These methods include homomorphic encryption [7-9], federated learning [10], differential privacy [10-14], model splitting [15] and blockchain [16].

Homomorphic encryption also known as implicit homomorphism, which can perform calculations on the basis of ciphertext and the result obtained by decryption is consistent with the result directly using plaintext. Aono et al. [7] divided their model into the client side and server side, the server performs calculation on ciphertext, the client downloads the encrypted weights from the server, and then uses its private key to decrypt it. After that, the client uses the weights to train its local neural network, and iteratively optimizes the weights. The optimized weights are then encrypted and uploaded to the server. The above process is repeated until the weights cannot be optimized any more. Compared with [7], Zhu et al. [8] encrypted the input data and used the model trained on plaintext to make predictions. Meftah et al. [9] proposed a more efficient deep convolutional neural network model based on fully homomorphic encryption. The model has relatively less ciphertext expansion cost, and it uses an accumulator with lower circuit depth complexity to improve efficiency. Park et al. [10] proposed a more secure privacy-preserving federated learning model. In the model, each participant can use a different private key to encrypt local model parameters, and the server can update parameters based on a homomorphic mechanism.

Differential privacy models can effectively protect data privacy by adding noise without encryption. Shokri and Shmatikov [11] proposed and implemented a scheme that multiple parties learn the model by selective sharing of model parameters (i.e. the weights in neural network) but without sharing their input data and use differential privacy approaches to add noises in training phase. However, Aono et al. [7] pointed out that local information from each party may be leaked to honest-but-curious server in [11]. Then, Abadi et al. [12] proposed to achieve different levels of differential privacy by adding noises in the stage of weights updating. However, these methods suffer from a non-negligible degradation of model accuracy when high level of privacy protection is required. Fan et al. [13] protect privacy by directly adding noise to image. Phan et al. [14] achieved the purpose of differential privacy protection by adding adaption Laplacian noise to the features that are less related to the results of the deep learning model.

In 2019, Yu et al. [15] divided the first convolutional layer of the CNN model into local sites and the remaining part into the server. The first part in local site uses step-wise activation function instead of the original activation function to defend the attacker. Lyu et al. [16] proposed a decentralized privacy-preserving deep learning framework with fairness considerations. The framework employs the blockchain technology to construct a decentralized structure for fair and privacy-preserving deep learning, and it uses differential privacy model to prevent the leakage of privacy in local training data.

\subsection{Negative Database}
The formal definition of negative database is given in [3]. Suppose all the strings in a database (\textit{DB}) are \textit{m}-bit, and the universal set of all \textit{m}-bit strings is denoted as U. The complementary set is denoted as U-\textit{DB}. Usually, the size of U-\textit{DB} is much larger than \textit{DB} so that it would be impractical to store all the strings in U-\textit{DB} explicitly. Therefore, we need to compress it and get a compact form named \textit{NDB} by introducing a “don’t-care” notation ‘*’, which can represent both ‘0’ and ‘1’. Generally, the bit with ‘0’ or ‘1’ is called as specific bit in NDBs and that with ‘*’ as unspecific bit. An example of NDB is shown in Table 1.

\begin{table}
	\centering
	\caption{\centering{An example 	of negative database}}
	\footnotesize
	\setlength{\tabcolsep}{10mm}
	\begin{tabular}{cccc}
	    \\
 		\toprule 
		\textit{DB}& \textit{U}–\textit{DB}&$\textit{NDB}_{1}$&$\textit{NDB}_{2}$ \\ 
		\midrule 
		000   &   001   &   0*1   &   **1    \\
        101   &   011   &   1*0   &   1*0\\
        010   &   100   &         & \\
              &   110   &      &\\
              &   111   &      &\\
 		\bottomrule 
	\end{tabular}
	\vspace{-3mm}
\end{table}

Negative database has many applications in the field of privacy-preserving. In the field of password authentication, Zhao et al. [17] first proposed an one-time password authentication scheme based on \textit{NDB}. The authentication data is converted to \textit{NDB} before being transmitted to the network, and the original data cannot be recovered even if the data is stolen by attacker. Bringer and Chabanne [18] proposed to use a negative database for biometrics. Subsequently, Zhao et al. [19] proposed a negative iris recognition scheme, which constructed a secure iris recognition method using a negative database generation algorithm and analyzed the security and effectiveness of the scheme. Zhao et al. [20] proposed NDBIris-\uppercase\expandafter{\romannumeral2} to protect iris data. The privacy data is firstly transformed irreversibly by local ranking technology, and the processed data is then converted into a negative database for identification. Liu et al. [21] applied the negative database to the K-means clustering algorithm to protect the privacy of data. Hu et al. [22] applied the \textit{K}-hidden algorithm to the kmeans clustering algorithm and proposed a Euclidean distance metric for negative databases. Zhu et al. [23] proposed a secure three-party authentication cross-protocol based on a negative database. The third-party server only plays the role of authentication and auxiliary transmission, and at the same time, the private information of client will not be leaked to each other.

\section{Preliminaries}
\subsection{\textit{QK}-hidden algorithm}
Zhao et al. [4] proposed the \textit{QK}-hidden algorithm and applied it to privacy-preserving \textit{K}-means clustering algorithm. In this paper, we use the \textit{QK}-hidden algorithm to generate \textit{NDB}s because it can control the utility of generated \textit{NDB}s granularly. The pseudo code of the \textit{QK}-hidden algorithm is shown in algorithm 1.

\begin{algorithm}[htb] 
\small
\caption{ The \textit{QK}-hidden algorithm [4]} 
\label{alg:Framwork} 
{\textbf{Input:}} an \textit{m}-bit string \textit{s}; length of attributes \textit{L}; a constant \textit{r};
the probability parameters $\textit{p}=\{\textit{p}_{1} {\dots} \textit{p}_{\textit{K}}\} and \textit{Q}=\{\textit{q}_{1} {\dots} \textit{q}_{\textit{L}}\}$ \\
{\textbf{Output:}} $\textit{NDB}_{\textit{s}}$
\begin{algorithmic}[1] 
\STATE $\textit{NDB}_{\textit{s}}$ ${\gets}$ ${\emptyset}$   
\label{ code:fram:extract }
\STATE \textit{N} ${\gets}$ \textit{m} ${\times}$ \textit{r} 
\label{code:fram:trainbase}
\STATE Initialize \{\textit{P}$_{0}$, \textit{P}$_{1}$, ${\dots}$ , \textit{P}$_{\textit{\textit{K}}}$\} : \textit{P}$_{0}$ ${\gets}$ 0, \textit{P}$_{\textit{i}}$ ${\gets}$ \textit{p}$_{1}$ + ${\dots}$ + \textit{p}$_{\textit{i}}$
\label{code:fram:add}
\WHILE{|$\textit{NDB}_{\textit{s}}$| < \textit{N}}
\label{code:fram:classify}
\STATE   Initialize a record ${\tau}$ as a string of \textit{m} unspecific bits ‘*’
\label{code:fram:select}
\STATE    \textit{rndp} ${\gets}$ \textit{random}([0,1))
\label{code:fram:select}
\STATE    Find the \textit{i} that satisfies:  \textit{P}$_{\textit{i}-1}$ ${\leq}$ \textit{rndp} < \textit{P}$_{\textit{i}}$
\label{code:fram:select}
\STATE   Randomly select \textit{i} bits of ${\tau}$ to be different from \textit{s} by probabilities \textit{Q}
\label{code:fram:select}
\STATE    Randomly select remaining \textit{K} ${-}$ \textit{i} bits of ${\tau}$ to be the same with \textit{s}
\label{code:fram:select}
\STATE    $\textit{NDB}_{\textit{s}}$ ${\gets}$  $\textit{NDB}_{\textit{s}}$ ${\cup}$ ${\tau}$
\label{code:fram:select}
\ENDWHILE
\label{code:fram:select}
\RETURN $\textit{NDB}_{\textit{s}}$ 
\end{algorithmic}
\end{algorithm}

The input of \textit{QK}-hidden algorithm is an \textit{m}-bit private binary string \textit{s} (hidden string), length of attributes \textit{L}, the number of attributes \textit{M}  (\textit{m} = \textit{L} × \textit{M}), a parameter \textit{r} (used for controlling the size of \textit{NDB}s), \{\textit{p}$_{1}$ {\dots} \textit{p}$_{\textit{K}}$\} (probability parameters for controlling the distribution of \textit{K} different types of records) and \{\textit{q}$_{1}$ …\textit{q}$_{\textit{L}}$\}(used for controlling the probabilities of selecting each bit in an attribute to be different from \textit{s}, and \textit{q}$_{1}$ + {\dots} + \textit{q}$_{\textit{L}}$ = 1). The main steps of \textit{QK}-hidden algorithm are: First, in steps 1-3, \textit{NDB}s is initialized as the empty set, \textit{N} is set to \textit{m} × \textit{r}, and the temporary variable \textit{P}$_{\textit{i}}$ (\textit{i}=1…\textit{K}) is assigned to the sum of \textit{p}$_{1}$…\textit{p}$_{\textit{i}}$; then, in steps 4-11, \textit{N} iterations are performed and one record is generated at each iteration. Specifically, a record \textit{${\tau}$} is initialized as a string of \textit{m} ‘*’s. Then, a random value is generated in step 6 and if it locates in the interval [\textit{P}$_{\textit{i}-1}$, \textit{P}$_{\textit{i}}$), then \textit{${\tau}$} will be generated as a type \textit{i} record. Generally, type \textit{i} record is generated with the probability \textit{p}$_{\textit{i}}$ because the length of [\textit{P}$_{\textit{i}-1}$, \textit{P}$_{\textit{i}}$) is \textit{p}$_{\textit{i}}$. When generating the type \textit{i} record \textit{${\tau}$}, \textit{i} bits of \textit{${\tau}$} are randomly selected to be different from the hidden string \textit{s} according to parameters \{\textit{q}$_{1}$ {\dots} \textit{q}$_{\textit{L}}$\}, and other \textit{K} ${-}$ \textit{i} bits of \textit{${\tau}$} are randomly selected to be the same to \textit{s}. Finally, \textit{N} records are generated and added to \textit{NDB}s, and \textit{NDB}s is returned as output in step 12. Note that, the process of selecting each of the \textit{i} bits of \textit{${\tau}$} in step 8 is similar to process of steps 6-7. In step 8, a random number \textit{j} will be selected from [1, \textit{L}] with the probability \textit{q}$_{\textit{j}}$, where \textit{L} is the length of attributes or variables. Then, an attribute of \textit{${\tau}$} is randomly chosen and its \textit{j}th bit is selected to be different from \textit{s}. This is also the difference between the \textit{QK}-hidden algorithm and the \textit{K}-hidden algorithm. The \textit{K}-hidden algorithm can be regarded as a special case of the \textit{QK}-hidden algorithm with \textit{q}$_{1}$ = {\dots} = \textit{q}$_{\textit{L}}$ = 1/\textit{L}.

\section{NegDL}
In this section, we propose a method named NegDL based on \textit{NDB} to preserve data privacy during deep learning. The NegDL is mainly divided into two parts, and its framework is shown in Fig. 1. First, the private dataset \textit{DB} = \{{\textit{X}$_{1}$ … \textit{X}$_{\textit{n}}$}\} will be converted to a set of negative databases \textit{NDB} = \{{\textit{NDB}$_{1}$  ... \textit{NDB}$_{\textit{n}}$}\} by the \textit{QK}-hidden algorithm, where \textit{NDB}$_{\textit{i}}$ (\textit{i}=1…\textit{n}) is the negative database generated from the private data \textit{X}$_{\textit{i}}$, which has \textit{M} attributes. Then the sketches \textit{S} = \{{\textit{S}$_{1}$ … \textit{S}$_{\textit{n}}$}\} from the negative databases in \textit{NDB} will be extracted, where \textit{S}$_{\textit{i}}$ is the sketch of \textit{NDB}$_{\textit{i}}$. Second, the sketches in \textit{S} instead of the raw data in \textit{DB} will be used as the input to the neural network for deep learning. Therefore, the data owner can convert personal data to negative databases before transmitting to the server conducting the process of deep learning, and thus, the data privacy will not be disclosed.

\begin{figure*}[tbp]
	\centerline{\includegraphics[width=4.66in]{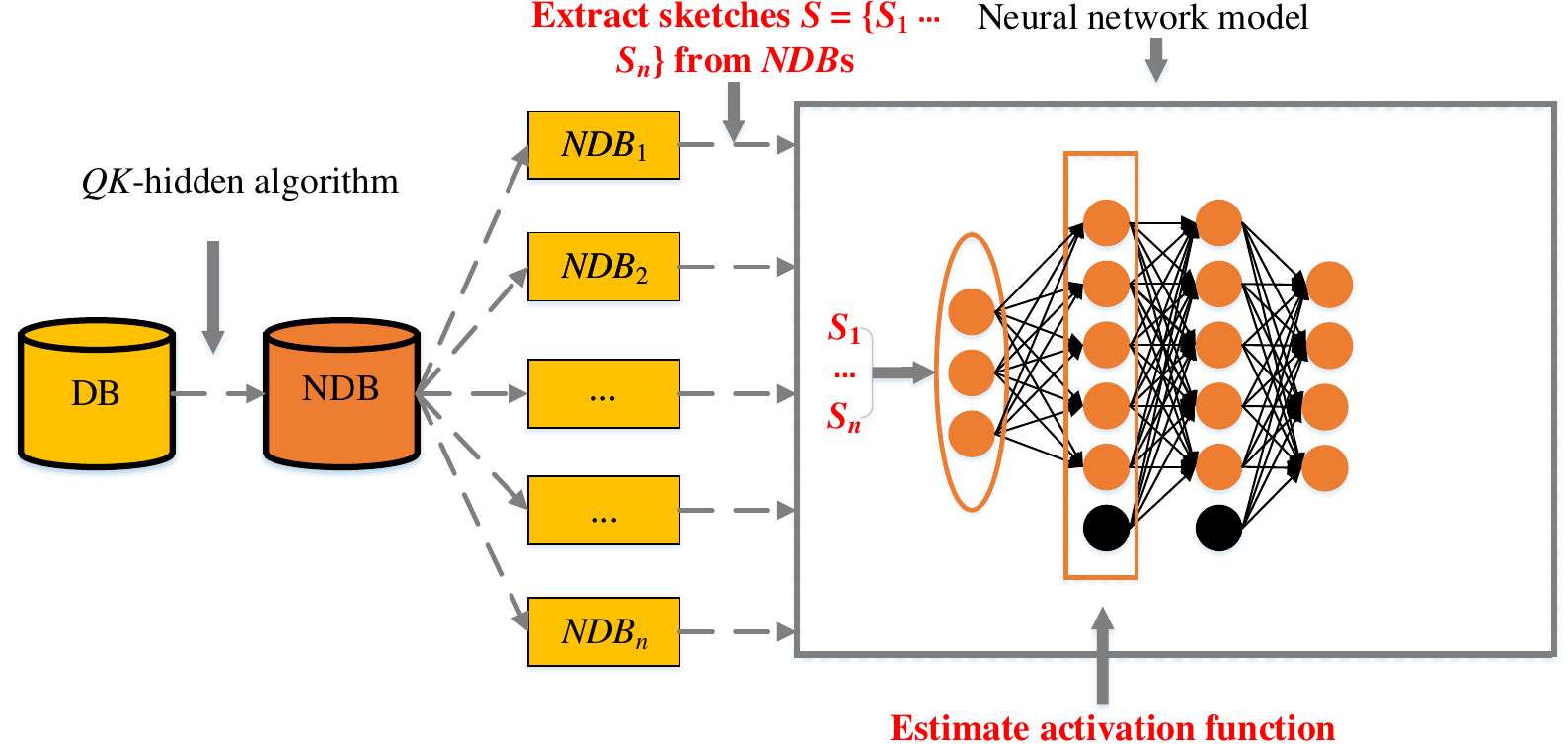}}
	\caption{\centering The overview of NegDL.}
	\label{fig1}
\end{figure*}
\subsection{Extracting  Sketches}
Specifically, in this paper, a sketch is a two-dimensional array, which compresses the \textit{NDB} records and retains necessary information for learning and inference. The sketches of negative databases are extracted by algorithm 2. The input is a negative database \textit{NDB}$_{\textit{i}}$ generated from the private data \textit{X}$_{\textit{i}}$ by the \textit{QK}-hidden algorithm, and the output is the sketch \textit{S}$_{\textit{i}}$ extracted from \textit{NDB}$_{\textit{i}}$. \textit{S}$_{\textit{i}}$ is a 2-dimensional array with 2\textit{m} numbers. For \textit{j} = 1 ... \textit{m} (\textit{m} is the length of \textit{X}$_{\textit{i}}$), \textit{S}$_{\textit{i}}$[\textit{j}][0] denotes the number of ‘0’s at the \textit{j}-th bit of all the records in \textit{NDB}$_{\textit{i}}$, and \textit{S}$_{\textit{i}}$[\textit{j}][1] denotes the number of ‘1’s at this bit.

\begin{algorithm}[htb] 
\small
\caption{ Generating the sketch from an \textit{NDB}.} 
\label{alg:Framwork} 
{\textbf{Input:}} $\textit{NDB}_{\textit{i}}$\\
{\textbf{Output:}} $\textit{S}_{\textit{i}}$
\begin{algorithmic}[1] 
\STATE Initialize \textit{S}$_{\textit{i}}$ = \{0\}  
\label{ code:fram:extract }
\STATE \textbf{for} each record ${\tau}$ in \textit{NDB}$_{\textit{i}}$ \textbf{do}
\label{code:fram:trainbase}
\STATE \qquad \textbf{for} each specific bit \textit{j} in ${\tau}$ \textbf{do} 
\label{code:fram:add}
\STATE  \qquad \qquad \textbf{if} the value of ${\textit{j}}$ is 1
\label{code:fram:select}
\STATE    \qquad \qquad \qquad \textit{S}$_{\textit{i}}$[\textit{j}][1]+=1 
\label{code:fram:select}
\STATE    \qquad \qquad \textbf{else}
\label{code:fram:select}
\STATE    \qquad \qquad \qquad \textit{S}$_{\textit{i}}$[\textit{j}][0]+=1 
\label{code:fram:select}
\STATE   \qquad \textbf{end for}
\label{code:fram:select}
\STATE   \textbf{end for}
\label{code:fram:select}
\RETURN $\textit{S}_{\textit{i}}$ 
\end{algorithmic}
\end{algorithm}

In step 1, \textit{S}$_{\textit{i}}$[\textit{j}][0] and \textit{S}$_{\textit{i}}$[\textit{j}][1] for \textit{j} = 1...\textit{m} are initialized to 0. At steps 2-9, the specific bits in \textit{NDB}$_{\textit{i}}$ are enumerated and the number of ‘0’s at the \textit{j}-th bit of all records in \textit{NDB}$_{\textit{i}}$ is counted as \textit{S}$_{\textit{i}}$[\textit{j}][0] and the number of ‘1’s is counted as \textit{S}$_{\textit{i}}$[\textit{j}][1]. After that, \textit{S}$_{\textit{i}}$ is returned as output in step 10.

\subsection{Estimating Activation Function}
In NegDL, since the extracted sketches in \textit{S} are inputted to the neural network instead of original private data \textit{X}, we cannot calculate the activation function \textit{f}(\textit{z}) with \textit{X} anymore, and thus, we should estimate the result of \textit{f}(\textit{z}) according to the sketches. For Sigmoid, ReLU and tanh function, they are originally calculated as follows:
\begin{equation}
             f(z) = \frac{1}{1 + \textit{e$^{-\textit{z}}$}} = \frac{1}{1+e^{\tiny{
			\setlength{\arraycolsep}{0.2pt}
			-\left[
			\begin{array}{ccc}
			x_1 & \cdots & x_M
			\end{array}
			\right]
			\left[
			\begin{array}{rrr}
			w_1 \\
	    	\vdots \\
			w_M \\
			\end{array}\right]}}}
\end{equation}

\begin{equation}
             f(z) = max({0, \textit{z}}) = max(0,{\tiny{
			\setlength{\arraycolsep}{0.2pt}
			\left[
			\begin{array}{ccc}
			x_1 & \cdots & x_M
			\end{array}
			\right]
			\left[
			\begin{array}{rrr}
			w_1 \\
			\vdots \\
			w_M \\
			\end{array}\right]}})
\end{equation}

\begin{equation}
\setlength{\arraycolsep}{0.2em}
f(z) = \frac{e^z-e^{-z}}{e^z+e^{-z}}=\frac{e^{2\tiny\left[\begin{array}{ccc}
		x_1 & \cdots & x_M
		\end{array}\right]\left[\begin{array}{rrr}
		w_1 \\
		\vdots \\
		w_M \\
		\end{array}\right]}-1}{e^{2\tiny\left[\begin{array}{ccc}
		x_1 & \cdots & x_M
		\end{array}\right]\left[\begin{array}{rrr}
		w_1 \\
		\vdots \\
		w_M \\
		\end{array}\right]}+1}
\end{equation}

In the above formula, \textit{z} represents the result of linear transformation and can be expressed as               $\small
			\setlength \arraycolsep{0.15em}
			\left[
			\begin{array}{ccc}
			x_1 & \cdots & x_M
			\end{array}
			\right]
			\left[
			\begin{array}{rrr}
			w_1 \\
			\vdots \\
			w_M \\
			\end{array}\right]$
for a certain neuron, where \textit{x} = \textit{x}$_{1}$ … \textit{x}$_{\textit{M}}$ (\textit{x} ${\in}$ \textit{X}) denotes the original private input of neural network. The process is described in Fig. 2. In NegDL, we should estimate the value of \textit{z} using sketches \textit{S} before calculating the activation function. Suppose the negative database of \textit{x} is \textit{NDB}$_{\textit{x}}$, then the probability that the \textit{i}-th bit of records in \textit{NDB}$_{\textit{x}}$ is different from \textit{x} can be calculated by the following formula [4].
\begin{equation}
P_{diff}\left[i\right]=\frac{\sum_{j=1}^{K}j\times p_j\times q_i}{\sum_{j=1}^{K}j\times p_j\times q_i+\sum_{j=1}^{K}(K-j)\times p_j\times\frac{1}{L}}
\end{equation}

\begin{figure*}[tbp]
	\centerline{\includegraphics[width=5.56in]{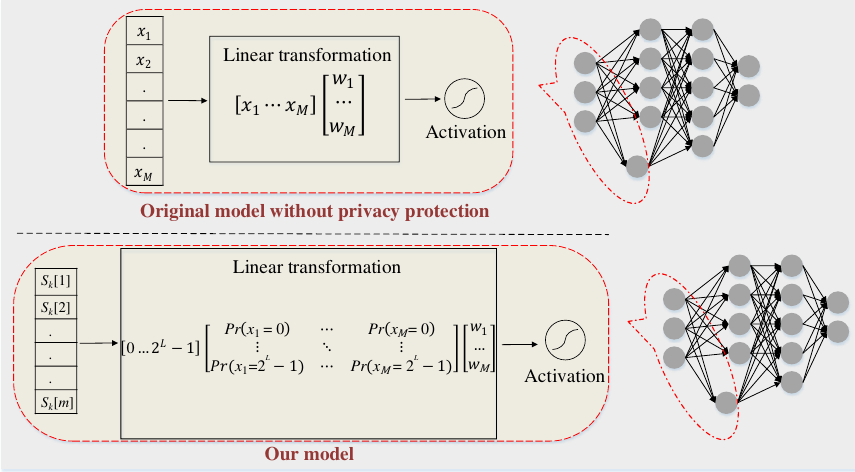}}
	\caption{\centering NegDL vs. original model without privacy protection.}
	\label{fig2}
\end{figure*}

Assume \textit{Pr}(\textit{x}$_{\textit{i}}^{\textit{j}}$= 0) is the probability that the \textit{j}-th (\textit{j}=1 {\dots} \textit{L}) bit of the \textit{i}-th (\textit{i}=1 {\dots} \textit{M}) attribute of \textit{x} is ‘0’ according to \textit{NDB}$_{\textit{x}}$, then it can be calculated by the following formula:
\begin{equation}
\small
\begin{aligned}
&Pr\left(x_i^j=0\right) = 
&\frac{\left(P_{same}\left[j\right]\right)^{n_0}\times \left(P_{diff}\left[j\right]\right)^{n_1}}{\left(P_{same}\left[j\right]\right)^{n_0}\times \left(P_{diff}\left[j\right]\right)^{n_1}+\left(P_{same}\left[j\right]\right)^{n_1}\times \left(P_{diff}\left[j\right]\right)^{n_0}}
\end{aligned}
\end{equation}

where \textit{n}$_{1}$ and \textit{n}$_{0}$ is the number of ‘1’s or ‘0’s at the \textit{j}-th bit of the \textit{i}-th attribute of all the records in \textit{NDB}$_{\textit{x}}$, respectively, which can be obtained from sketches \textit{S}, i.e. \textit{n}$_{0}$ = \textit{S}$_{\textit{k}}$[\textit{i} ${\times}$ \textit{L} + \textit{j}][0] and \textit{n}$_{1}$ = \textit{S}$_{\textit{k}}$[\textit{i} ${\times}$ \textit{L} + \textit{j}][1] if \textit{x} = \textit{X}$_{\textit{k}}$. \textit{Pr}(\textit{x}$_{\textit{i}}^{\textit{j}}$ = 1) = 1 ${-}$ \textit{Pr}(\textit{x}$_{\textit{i}}^{\textit{j}}$ = 0). After this, we can calculate the probability that the value of the \textit{i}-th attribute of \textit{x} is \textit{d}(0 ${\leq}$ \textit{d} ${\leq}$ 2$^{\textit{L}}$-1 ) by:
\begin{equation}
Pr\left(x_i=d\right)=Pr\left(x_i^{bin}=b_1\dots b_L\right)=\prod_{j=1}^{L}Pr\left(x_i^j=b_j\right)
\end{equation}
where \textit{x}$_{\textit{i}}^{\textit{bin}}$  is the binary representation of $\textit{x}_{\textit{i}}$ and the binary representation of \textit{d} is \textit{d}$^{\textit{bin}}$ = \textit{b}$_{1} {\cdots}  \textit{b}_{L}$. Therefore, we have:
\begin{equation}
\setlength{\arraycolsep}{0.2em}
\begin{aligned}
&\overline{\textit{z}}=\left[\begin{array}{ccc}
0 & \cdots & 2^L-1
\end{array}\right] 
&\left[\begin{array}{ccc}
Pr\left(x_1=0\right) & \cdots & Pr\left(x_M=0\right) \\
\vdots & \ddots & \vdots \\
Pr\left(x_1=2^L-1\right) & \cdots & Pr\left(x_M=2^L-1\right)\\
\end{array}\right]\left[\begin{array}{r}
w_1 \\
\vdots \\
w_M \\
\end{array}\right]
\end{aligned}
\end{equation}

We can use the estimated value of ${\overline{\textit{z}}}$ in (7) to substitute \textit{z} in formula (1)-(3) for estimating the result of the activation function \textit{f}(\textit{z}). The process of estimating the result of the activation function from sketches in NegDL is shown in Fig. 2. After that, the result of the activation function will be inputted to the next layer until the end of forward propagation. 

\begin{algorithm*}[htb] 
\small
\caption{ NegDL} 
\label{alg:Framwork} 
{\textbf{Training phase:}}\\ 
{\textbf{Input:}} 
The raw data \textit{X} = \{{\textit{X}$_{1}$ ${\dots}$ \textit{X}$_{\textit{n}}$}\} and labels \textit{y} = \{{\textit{y}$_{1}$ ${\dots}$ \textit{y}$_{\textit{n}}$}\}\\
\hspace*{0.95cm} Learning rate ${\eta}$, loss function \textit{Loss}((\textit{W}, \textit{X}$_{\textit{i}}$), \textit{y}$_{\textit{i}}$)\\
\hspace*{0.95cm} Maximum number of iterations \textit{E}, batch size \textit{t} \\
{\textbf{Output:}}
\textit{W} = \{{\textit{W}$_{1}$ ${\dots}$ \textit{W}$_{\textit{n}}$}\}
\begin{algorithmic}
\STATE {$\textbf{User:}$} \\       
\label{ code:fram:extract }
\STATE1. Convert \textit{X}=\{{\textit{X}$_{1}$ ${\dots}$ \textit{X}$_{\textit{n}}$}\} to negative databases \textit{NDB} =\{{\textit{NDB}$_{1}$ ${\dots}$ \textit{NDB}$_{\textit{n}}$ }\} by the \textit{QK}-algorithm
\label{code:fram:select}
\STATE2. Extract sketch \textit{S}$_{\textit{i}}$ (\textit{i} = 1 ${\dots}$ \textit{n}) from \textit{NDB}$_{\textit{i}}$ by algorithm 2 and then send \textit{S} to the server
\STATE \textbf{Server:}
\label{code:fram:select}
\STATE3. Receive \textit{S} from all users and input them to the neural network
\label{code:fram:select}
\STATE   4. Initialize weight \textit{W} = \{{\textit{W}$_{1}$ ${\dots}$ \textit{W}$_{\textit{n}}$}\} randomly
\label{code:fram:select}
\STATE    5. \textbf{while}(number of iterations ${\leq}$ \textit{E} or \textit{Loss} not reaches minimum)
\label{code:fram:select}
\STATE    6. \qquad Randomly select ${\omega}$ users and use their sketches \textit{S} for forward propagation
\label{code:fram:select}
\STATE   7. \qquad  Estimate \textit{Loss}((\textit{W}, \textit{X}$_{\textit{i}}$), \textit{y}$_{\textit{i}}$)) using \textit{S} of ${\omega}$ users based on (1)-(3) and (7)
\label{code:fram:select}
\STATE    8. \qquad Compute gradients${:{\Tilde{g}} = \nabla_{\textit{W}}\frac{1}{\textit{t}}\sum_{\textit{i} = 1} ^\textit{t}\textit{Loss}((\textit{W},\textit{X}_{i}),\textit{y}_{i})
\label{code:fram:select}}$
\STATE    9. \qquad Update weights: $\textit{W} = \textit{W} - \eta\Tilde{g}$
\label{code:fram:select}
\STATE  10. \textbf{end while}
\label{code:fram:select}
\STATE  11. \textbf{return} \textit{W}
\end{algorithmic} \rule{17.0cm}{0.05em}
{\textbf{Testing phase:}}\\ 
{\textbf{Input:}} The raw data \textit{X}$_{\textit{test}}$ \\
{\textbf{Output:}} Predict result \textit{y}$_{\textit{test}}$ 
\begin{algorithmic}
\STATE \textbf{User:}       
\label{ code:fram:extract }
\STATE 1. Convert testing data \textit{X}$_{\textit{test}}$ to negative databases \textit{NDB}$_{\textit{test}}$ by algorithm 1      
\label{ code:fram:extract }
\STATE 2. Extract sketch \textit{S}$_{\textit{test}}$ from \textit{NDB}$_{\textit{test}}$ by algorithm 2 and then send \textit{S}$_{\textit{test}}$ to the server
\label{ code:fram:extract }
\STATE \textbf{Server:}       
\label{ code:fram:extract }
\STATE 3. Receive \textit{S}$_{\textit{test}}$ from all users and input them to the neural network obtained in training process      
\label{ code:fram:extract }
\STATE 4. Return inference/predict result to users       
\label{ code:fram:extract }
\end{algorithmic}
\end{algorithm*}

\subsection{Algorithm Description}
The details of NegDL are shown in algorithm 3. In the training phase, the input of NegDL is the private data \textit{X} = \{{\textit{X}$_{1}$ ${\cdots}$ \textit{X}$_{\textit{n}}$}\} and labels \textit{y} = \{{\textit{y}$_{1}$ ${\cdots}$ \textit{y}$_{\textit{n}}$}\}, the learning rate ${\eta}$, the loss function \textit{Loss}, the maximum number of iterations \textit{E} and batch size \textit{t}. The output is the weights \textit{W}. This algorithm is divided into a user part and a server part. In steps 1-2, the user converts his data \textit{X} into \textit{NDB}s by the \textit{QK}-hidden algorithm and extracts sketches \textit{S} from \textit{NDB}s, and then sends \textit{S} to the server for conducting the training process of deep learning. After receiving the sketches \textit{S} from all users, the server initializes weights \textit{W} randomly. In steps 5-9, the server iteratively updates \textit{W} until the number of iterations reaches \textit{E} or \textit{Loss}((\textit{W}, \textit{X}$_{\textit{i}}$), \textit{y}$_{\textit{i}}$) reaches the minimum. In the testing phase, similar to the training phase, the user first converts the testing data to \textit{NDB}s and submits them to the server. Next, the server estimates the result of the activation function according to the sketches by formulas (1)-(3) and (7), and then, it uses the model obtained in the training phase to predict the data and returns the result to the user.

\subsection{Complexity Analysis}

At the user side in NegDL, the computational complexity of step 1 is O(\textit{m} ${\times}$ \textit{n}) (\textit{K} and \textit{r} are regarded as constants). It takes O(\textit{m} ${\times}$ \textit{n}) computations to extract sketches from \textit{NDB}s in step 2. For the user in the testing phase, the computational complexity of step 1 and step 2 is O(\textit{m}) if only one data is tested. Overall, the computational complexity for a user is O(\textit{m} ${\times}$ \textit{n}) in training phase and O(\textit{m}) in testing phase, which is the same as the original deep learning algorithm without privacy protection.

At the server side in NegDL, as the size of sketches is two times to the size of original data, the complexity of step 3 is also O(\textit{m} ${\times}$ \textit{n}). In step 7, it takes O(\textit{n} ${\times}$ \textit{M} ${\times}$ 2$^{\textit{L}}$) computations to estimate the activation function in (7), and we usually regard \textit{L} as a constant, thus the complexity of this step is equivalent to the complexity of the original algorithm without privacy protection. Steps 4-6 and 8-10 are same to the original algorithm except using sketches instead of original data, so the complexity also keeps unchanged. In testing phase, since the network trained in NegDL has the same size as network trained in original algorithm, the computational complexity of them is also the same.

Overall, according to the above analysis, the proposed approach has the same computational complexity as the original model without privacy protection in both training phase (if the algorithm is terminated according to “number of iterations \textit{E}”) and the testing phase.

\section{Experiments}
In this section, we conduct several groups of experiments on two benchmark datasets, i.e. MNIST and CIFAR-10, to validate the effectiveness of NegDL. For the \textit{QK}-hidden algorithm, we set \textit{K} = 3, \textit{p}$_{1}$ = 0.70, \textit{p}$_{2}$ = 0.24, \textit{p}$_{3}$ = 0.06 and \textit{r} = 6.5 as default, which could generate \textit{NDB}s with balanced security and utility [21]. We implemented NegDL on a server with Intel Core CPU i5-8500, 24G RAM, and a Geforce RTX 3080Ti GPU.

\subsection{Experiment on MNIST}
MNIST [5] is a handwritten digit recognition dataset containing 60,000 training samples and 10,000 testing samples. Each sample in MNIST dataset consists of 28 ${\times}$ 28 pixels, and each pixel is represented by a gray value between 0 and 255. We use the LeNet-5 [28] model as the neural network architecture. The network architecture consists of two convolutional layers and two fully connected layers. All of convolutional layers use 5 ${\times}$ 5 kernels with stride 1, followed by a ReLU and 2 ${\times}$ 2 max pools. The two fully connected layers contain 120 and 84 neurons, respectively, and finally the output layer. In the training phase, we set the learning rate to $10^3$, the batch size to 128, and adopt the cross-entropy loss function and Adam optimizer. The training process contains 100 epochs. In the model without privacy protection, the testing accuracy reaches 99.07\%, as shown in Fig. 3.

\begin{table*}
	\centering
	
	\caption{\centering The twelve groups of parameters \textit{Q} in MNIST and CIFAR-10.}
	\footnotesize
	\begin{tabular}{lcccccccccc}
 		\toprule  
		 &\textit{q}$_{1}$&\textit{q}$_{2}$&\textit{q}$_{3}$&\textit{q}$_{4}$&\textit{q}$_{5}$&\textit{q}$_{6}$&\textit{q}$_{7}$&\textit{q}$_{8}$&\textit{P}$_{\textit{diff}}[1]$&\textit{P}$_{\textit{diff}}[8]$\\ 
		\midrule 
		\textit{Q}$_{1}$&0.05&0.10&0.10&0.10&0.10&0.10&0.10&0.35&0.25&0.70\\
        \textit{Q}$_{2}$&0.10&0.10&0.10&0.10&0.10&0.10&0.10&0.30&0.40&0.67 \\
        \textit{Q}$_{3}$&0.125&0.125&0.125&0.125&0.125&0.125&0.125&0.125&0.45&0.45\\
        \textit{Q}$_{4}$&0.20&0.10&0.10&0.10&0.10&0.10&0.10&0.20&0.57&0.57\\
        \textit{Q}$_{5}$&0.30&0.10&0.10&0.10&0.10&0.10&0.10&0.10&0.67&0.40\\
        \textit{Q}$_{6}$&0.40&0.05&0.05&0.05&0.05&0.05&0.05&0.30&0.73&0.67 \\
        \textit{Q}$_{7}$&0.50&0.05&0.05&0.05&0.05&0.05&0.05&0.20&0.77&0.57 \\
        \textit{Q}$_{8}$&0.60&0.05&0.05&0.05&0.05&0.05&0.05&0.10&0.80&0.40 \\
        \textit{Q}$_{9}$&0.70&0&0.05&0.05&0.05&0.05&0.05&0.05&0.82&0.25 \\
        \textit{Q}$_{10}$&0.80&0&0&0&0.05&0.05&0.05&0.05&0.84&0.25 \\
        \textit{Q}$_{11}$&0.90&0&0&0&0&0&0.05&0.05&0.85&0.25 \\
        \textit{Q}$_{12}$&0.95&0&0&0&0&0&0&0.05&0.86&0.25 \\
 		\bottomrule  
	\end{tabular}
	\vspace{-3mm}
\end{table*}

\begin{figure*}[tbp]
	\centerline{\includegraphics[width=3.56in]{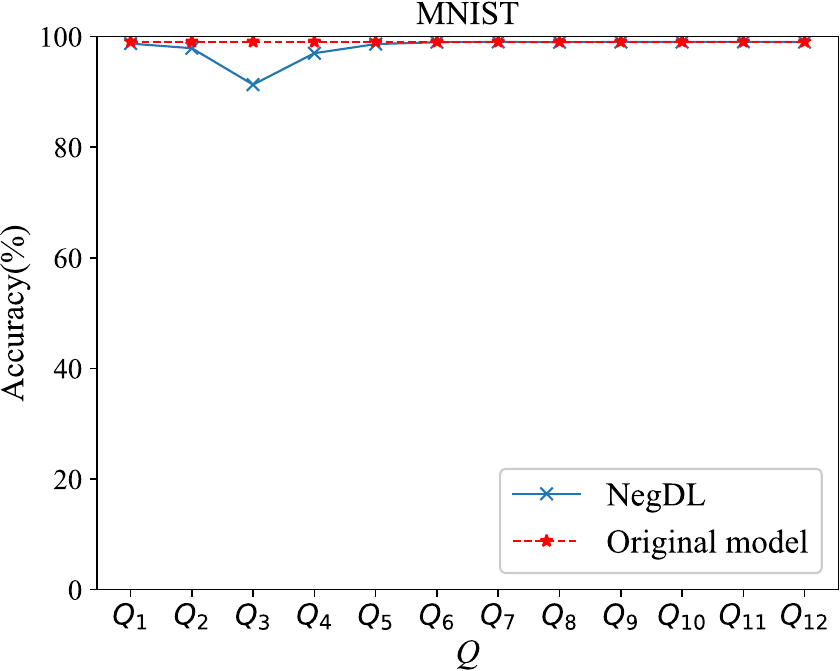}}
	\caption{\centering The impact of \textit{q} on accuracy on MNIST.}
	\label{fig3}
\end{figure*}

\begin{figure*}[tbp]
	\centerline{\includegraphics[width=3.56in]{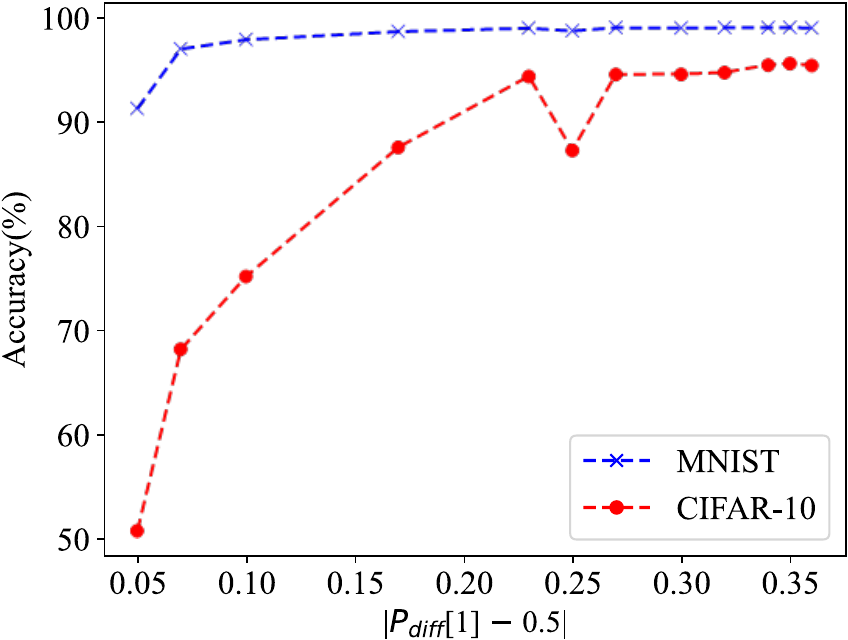}}
	\caption{ \centering  Accuracy for different $|\textit{P}_{\textit{diff}}[1] - 0.5|$ on MNIST and CIFAR-10 datasets.}
	\label{fig4}
\end{figure*}

For NegDL, we first represent each pixel as an attribute with an 8-bit binary string, and the bit is more significant from right to left. For example, if the data is (16, 59, 43, 8), the first attribute can be converted to the binary string “00010000”, and finally, we concatenate binary strings of 4 attributes into a long string “00010000001110110010101100001000”. As each instance contains 784 attributes, each sample is finally expressed as a binary string of 28 ${\times}$ 28 ${\times}$ 8 = 6272 bits. In NegDL, we mainly investigate the impact of \textit{q}$_{1}$. Twelve different parameter settings \textit{Q}$_{1}$ ${\cdots}$ \textit{Q}$_{12}$ are used and shown in Table 2. In these settings, we change the value of \textit{q}$_{1}$ from 0.05 to 0.95, and fix \textit{q}$_{\textit{i}}$ (\textit{i} = 2 ${\cdots}$ 7) to 0, 0.05 or 0.10. \textit{q}$_{8}$ is set to 1 – \textit{q}$_{1}$ – ${\cdots}$ – \textit{q}$_{7}$. Fig. 3 shows the accuracy of NegDL and the original model (without privacy-preserving). The accuracy of NegDL is 98.71\%, 97.88\%, 91.31\%, 96.99\%, 98.64\%, 98.97\%, 99.02\%, 98.97\%, 99.02\%, 99.01\%, 99.04\% and 98.98\% for \textit{Q}$_{1}$, ${\cdots}$, \textit{Q}$_{12}$, respectively. It is worth noting that the accuracy decreases sharply to 91.31\% when \textit{Q}$_{3}$ is used. This is because \textit{q}$_{1}$ ${\cdots}$ \textit{q}$_{8}$ are all set to 0.125 in \textit{Q}$_{3}$, and |\textit{P}$_{\textit{diff}}$[1] ${-}$ 0.5| is 0.05, which is the smallest among the twelve \textit{Q}s, so the accuracy is the lowest. Similarly, Fig. 4 shows that the accuracy roughly increases with the increase of |\textit{P}$_{\textit{diff}}$[1] ${-}$ 0.5|. In particular, when \textit{q}$_{1}$ ${\geq}$ 0.4, the accuracy is not less than 98.97\% and close to the accuracy of the model without privacy protection. By comparing \textit{Q}$_{1}$ ${\cdots}$ \textit{Q}$_{5}$ with \textit{Q}$_{6}$ ${\cdots}$ \textit{Q}$_{12}$, we find another phenomenon that changing the bit which is more significant would take more effect on the testing accuracy. Moreover, we observe that the accuracy is difficult to improve anymore from \textit{Q}$_{8}$ to \textit{Q}$_{12}$, and sometimes it even decreases slightly. The reason is that the accuracy already reaches the peak value approximating to 99.07\% (the accuracy of the model without privacy protection), and the slight degradation might be caused by the randomness of the \textit{QK}-hidden algorithm.

\begin{figure*}[tbp]
	\centerline{\includegraphics[width=3.56in]{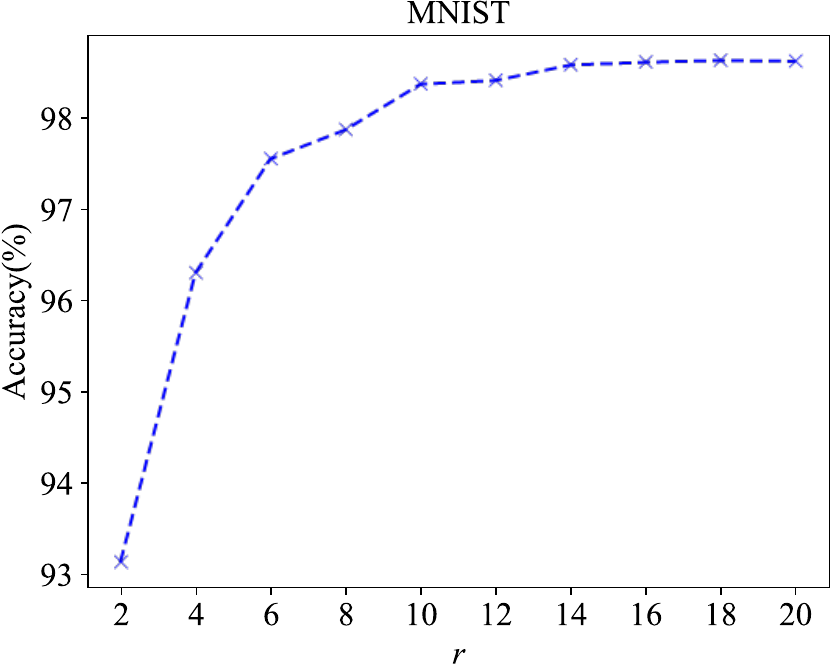}}
	\caption{\centering Fig. 5.	The impact of \textit{r} on accuracy on MNIST.}
	\label{fig5}
\end{figure*}

In addition, we also investigate the influence of parameter r on accuracy in experiments. We fix the parameters \textit{p} to (\textit{p}$_{1}$ = 0.70, \textit{p}$_{2}$ = 0.24, \textit{p}$_{3}$ = 0.06), \textit{K} to 3 and \textit{Q} to \textit{Q}$_{2}$, and vary \textit{r} from 2 to 20. Fig. 5 shows the relationship between r and accuracy. The result shows that with the increase of \textit{r}, the accuracy increases from 93.13\% to 98.63\%. The reason is that: as \textit{r} increases, more \textit{NDB} records are generated, and more information can be used for extracting sketches. Therefore, the data utility is higher. The accuracy tends to be stable when \textit{r} ${\geq}$  14, and the reason might be that the information for extracting sketches is already sufficient.

\subsection{Experiment on CIFAR-10}
CIFAR-10 [6] is a dataset with 10 categories (airplane, bird, cat, dog, etc.), which consists of 50,000 training samples and 10,000 testing samples. Each sample is a 3-channel color RGB image with 32 ${\times}$ 32 pixels.

\begin{figure*}[tbp]
	\centerline{\includegraphics[width=3.56in]{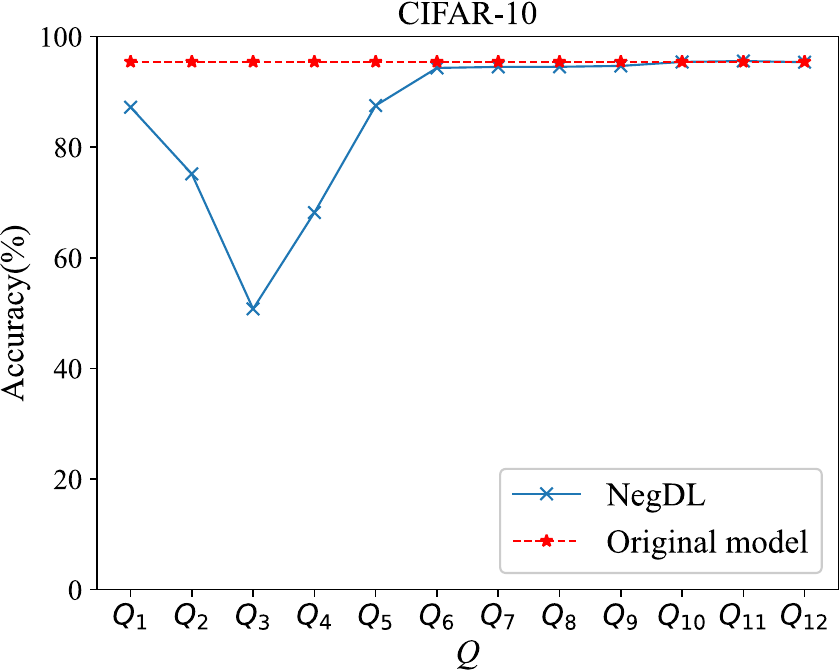}}
	\caption{\centering The impact of \textit{q} on accuracy on CIFAR-10.}
	\label{fig6}
\end{figure*}

We conduct the experiment with reference to the ResNet-50 [24] model. The first layer of the network is a convolutional layer with a filter size of 7 ${\times}$ 7 and contains 64 channels, followed by a max pooling layer. Then there are four residual blocks, containing 9, 12, 18 and 9 convolutional layers respectively. The features of the last convolutional layer are output to the average pooling layer and then passed to the fully connected layer, and finally output through softmax regression probabilities for each class. In particular, we also performs data augmentation processing on the input. Specifically, random cropping and horizontal flipping techniques are used for each input sample. We select the SGD optimizer and the cross-entropy loss function and set the batch size to 128 to train 300 epochs. The initial learning rate $\eta$ = 0.1, and use the CosineAnnealingLR learning rate adjustment strategy officially provided by pytorch to better train the model. In this strategy, the parameter T\_max is set to 300, and the learning rate is gradually reduced from 0.1 to 0.000003 after 300 epochs. The testing accuracy of the model without privacy protection reaches 95.44\% (as shown in Fig. 6).

For NegDL, each channel of the images in CIFAR-10 was converted into a binary string with 8192 bits. Like the experiment on MNIST, we choose the same settings for \textit{Q} in Table 2, i.e. \{\textit{Q}$_{1}$, ${\cdots}$ , \textit{Q}$_{12}$\}, and change \textit{q}$_{1}$ from 0.05 to 0.95. In Fig. 6, we achieve the accuracy 87.23\%, 75.15\%, 50.77\%, 68.17\%, 87.52\%, 94.32\%, 94.51\%, 94.52\%, 94.69\%, 95.41\%, 95.56\%,  95.37\% for \{\textit{Q}$_{1}$, ${\cdots}$ , \textit{Q}$_{12}$\}, respectively. Similarly, the accuracy for NegDL drops first and then rises with the change of \textit{q}$_{1}$. The worst accuracy is 50.77\%, which is also achieved with \textit{Q}$_{3}$ and \textit{q}$_{1}$ = ${\cdots}$ = \textit{q}$_{8}$ = 0.125. In contrast, when the parameters are set to \textit{Q}$_{11}$, the testing accuracy reaches 95.56\%, which is even slightly better than the accuracy without the privacy protection. By observing the results in Fig. 4, we find that the accuracy roughly increases with the increase of |\textit{P}$_{\textit{diff}}$[1] ${-}$ 0.5|. When \textit{q}$_{1}$ = 0.05 in \textit{Q}$_{1}$, |\textit{P}$_{\textit{diff}}$[1] ${-}$ 0.5| = 0.25, but the accuracy at this case is lower than the accuracy at \textit{q}$_{1}$ = 0.40 in \textit{Q}$_{6}$, this is because \textit{P}$_{\textit{diff}}$[2], ${\cdots}$ , \textit{P}$_{\textit{diff}}$[7] for \textit{Q}$_{1}$ are closer to 0.50.

\subsection{Utility  analysis}
In this part, we first introduce SSIM (structural similarity index measure) [25] and PSNR (peak signal-to-noise ratio) [26] before performing utility analysis. The formulas of SSIM and PSNR are as follows:

\begin{equation}
    SSIM(\textit{x},\textit{y}) = \frac{(2\mu_\textit{x}\mu_\textit{y} + \textit{c}_1)(2\sigma_{\textit{xy}}+\textit{c}_2)}{(\mu^2_\textit{x}+\mu^2_\textit{y}+\textit{c}_1)(\sigma^2_\textit{x}+\sigma^2_\textit{y}+\textit{c}_2)}
\end{equation}

where \textit{x} and \textit{y} represent two images, ${\mu}$ is the average, ${\sigma_\textit{x}^2}$ is the variance of \textit{x}, ${\sigma_{\textit{xy}}}$ is the covariance of \textit{x} and \textit{y}, $\textit{c}_1$ and $\textit{c}_2$ are constants.

\begin{equation}
             PSNR(\textit{x},\textit{y}) = 10 \cdot \log_{10}(\frac{\textit{MAX}_\textit{I}^2}{\textit{$MSE(x,y)$}})
\end{equation}

\begin{equation}
             MSE(\textit{x},\textit{y}) = \frac{1}{\textit{MN}} \sum_{\textit{i}=1}^{\textit{M}}\sum_{\textit{j}=1}^{\textit{N}}(\textit{x}_{\textit{ij}}-\textit{y}_{\textit{ij}})^2
\end{equation}

where $\textit{MAX}_\textit{I}$ is the maximum possible pixel value of the image. When using 8 bits images to represent pixels, $\textit{MAX}_\textit{I}$ is 255. \textit{MSE} represents mean squared error.

\begin{table*}[tb]
	\centering
	\setlength{\tabcolsep}{1mm}
	\caption{\centering{SSIM and PSNR for different \textit{Q}s.}}
	\footnotesize
	\begin{tabular}{lcccccccccccc}
 		\toprule  
	    Parameters & \textit{Q}$_{1}$ & \textit{Q}$_{2}$ & \textit{Q}$_{3}$ & \textit{Q}$_{4}$ & \textit{Q}$_{5}$ & \textit{Q}$_{6}$ & \textit{Q}$_{7}$ & \textit{Q}$_{8}$ & \textit{Q}$_{9}$ & \textit{Q}$_{10}$ & \textit{Q}$_{11}$ & \textit{Q}$_{12}$ \\ 
		\midrule 
		{PSNR}  (MNIST)&15.04&10.50&7.27&9.86&14.82&26.59&27.31&27.37&33.46&45.49&57.71&64.67 \\
		{SSIM}  (MNIST)&0.6278&0.4431&0.1645&0.3880&0.6220&0.8465&0.8494&0.8505&0.9189&0.9901&0.9993&0.9997 \\
        {PSNR}  (CIFAR-10)&18.19&14.14&12.87&13.57&18.04&27.14&27.94&27.98&33.93&45.42&56.07&61.73 \\
        {SSIM}  (CIFAR-10)&0.5663&0.3082&0.1660&0.2587&0.5576&0.8805&0.8941&0.8948&0.9648&0.9967&0.9987&0.9986 \\
 		\bottomrule  
	\end{tabular}
\end{table*}

The above two indicators are usually used to evaluate the quality of the image. The smaller value of SSIM implies the smaller similarity between  two images, and it also means that it is difficult for attackers to recover original image. Similarly, smaller values of PSNR indicates the worse the quality of the image. In this paper, we calculate SSIM and PSNR for twelve groups of different \textit{Q}s between the data that attackers guessed through sketches and the private data. Specifically, if an attacker steals sketch \textit{$S_x$}, for the \textit{i}-th attribute of private data \textit{x}, the attacker can guess the value using $\textit{d}={\sum_{\textit{j}=0}^{2^{\textit{L}}-1}\textit{j} \times Pr(x_i=j)}$. Where \textit{Pr}($x_i=j$) can be calculated by formula (6). The results in Table 3 show that the SSIM for \textit{Q}$_{3}$ on MNIST and CIFAR-10 are 0.1645 and 0.1660 respectively, and the PSNR are 7.27 and 12.87 respectively. This phenomenon indicates the images obtained by attackers have quite low utility. Moreover, The values of SSIM and PSNR increase with the change of \textit{Q}s, especially, the value of SSIM is close to 1 at \textit{Q}$_{10}$, \textit{Q}$_{11}$ and \textit{Q}$_{12}$. The result corresponds to the lowest privacy in previous experiments.

\subsection{Comparison}
In this part, we compare NegDL with the method for privacy-preserving deep learning based on differential privacy proposed by Abadi et al. [12].

\begin{table*}[tb]
\centering
	\caption{\centering{The  accuracy(\%) of [12] VS NegDL}}
	\footnotesize
	\begin{tabular}{cc|cc}
 		\toprule 
	    Parameter settings & MNIST & Parameter settings & CIFAR-10\\ 
		\midrule  
	    Abadi et al. [12](0.5, 10$^{-5}$) & 90 &  Abadi et al. [12](2, 10$^{-5}$) & 67\\
        Abadi et al. [12](2, 10$^{-5}$) & 95 &  Abadi et al. [12](4, 10$^{-5}$) & 70\\
        Abadi et al. [12](8, 10$^{-5}$) & \textbf{97} &  Abadi et al. [12](8, 10$^{-5}$) & \textbf{73}\\
        NegDL(\textit{Q}$_{1}$) & 98.11 & NegDL(\textit{Q}$_{1}$) & 61 \\
        NegDL(\textit{Q}$_{6}$) & 98.28 & NegDL(\textit{Q}$_{6}$) & 73 \\
        NegDL(\textit{Q}$_{12}$) & \textbf{98.46} & NegDL(\textit{Q}$_{12}$) & \textbf{77} \\
        Original model & \textbf{98.50} & Original model & \textbf{80} \\
 		\bottomrule 
	\end{tabular}
\end{table*}

In [12], they achieved the purpose of privacy preserving by adding noise to the gradient, we cannot directly calculate the SSIM and PSNR for it, so we use accuracy as an indicator to compare the two methods. Specifically, they used PCA projection to reduce the input data to 60 dimensions, and finally they achieved the testing accuracy 90\%, 95\%, and 97\% for different levels of privacy protection on MNIST. Similarly, we conducted experiments using the same neural network model. To be fair, we took three groups of parameters with the \textit{Q}$_{1}$, \textit{Q}$_{6}$, and \textit{Q}$_{12}$ accuracy respectively for comparison. The results in Table 4 show that NegDL could achieve better accuracy than [12] with three typical parameter settings. Moreover, NegDL gets an accuracy of 98.46\% with \textit{Q}$_{12}$, which is very close to the accuracy of the original model (without privacy protection). For CIFAR-10, Abadi et al. [12] also obtained three results (67\%, 70\%, 73\%) by adding large, medium and small noises, respectively, and they are shown in Table 4. The original model could achieve an accuracy of 80\% after 250 epochs of training with the architecture used in [12]. In our experiment, similarly, we present the results of NegDL with \textit{Q}$_{1}$, \textit{Q}$_{6}$, and \textit{Q}$_{12}$ parameter settings in Table 2. It is shown that NegDL achieves more than 5\% improvement on accuracy compared to the method of [12] when \textit{Q}$_{12}$ parameters is used. However, when the \textit{Q}$_{1}$ parameters are used, the accuracy decreases from 67\% to 61\%, and the reason might be the great precision loss on high-order bits as \textit{q}$_{2}$ ${\cdots}$ \textit{q}$_{7}$ are set to be 0.1 and large noises are produced on high-order bits. The experimental results show that our method is comparable to this method.

It is worth mentioning that NegDL can also achieve different accuracy by changing the parameters \textit{p} and \textit{q} in the \textit{QK}-hidden algorithm, and users could choose proper parameter settings according to their requirements in real-world applications. Generally, the setting of \textit{p} and \textit{q} corresponding to a larger |\textit{P}$_{\textit{diff}}$ – 0.5| usually results in a better accuracy and vice versa. 
\begin{table*}[tb]
	\centering
	\caption{\centering{\textit{G} for different \textit{Q}s.}}
	\small
	\begin{tabular}{lcccccccccccc}
 		\toprule  
	    Parameters & \textit{Q}$_{1}$ & \textit{Q}$_{2}$ & \textit{Q}$_{3}$ & \textit{Q}$_{4}$ & \textit{Q}$_{5}$ & \textit{Q}$_{6}$ & \textit{Q}$_{7}$ & \textit{Q}$_{8}$ & \textit{Q}$_{9}$ & \textit{Q}$_{10}$ & \textit{Q}$_{11}$ & \textit{Q}$_{12}$ \\ 
		\midrule 
		\textit{G}(MNIST)&1670&1936&3890&2295&1933&252&546&484&212&141&71&36 \\
        \textit{G}(CIFAR-10)&2177&2523&5082&2992&2526&331&716&631&278&183&91&47 \\
 		\bottomrule  
	\end{tabular}
\end{table*}

\subsection{Security analysis}
In this section, the security of NegDL is analyzed. First, it has been proven that reversing an \textit{NDB} to obtain its corresponding \textit{DB} is equivalent to solving a SAT formula, which is an \textit{NP}-hard problem [7]. Moreover, existing algorithms such as \textit{p}-hidden algorithm and \textit{K}-hidden can be used to generate hard-to-reverse \textit{NDB}s. In NegDL, we choose the \textit{QK}-hidden algorithm to generate negative databases, which is a general form of the \textit{K}-hidden algorithm. For attackers, the best way as far as we know to break the generated negative databases to disclose privacy is to guess the private data based on estimating model proposed in Section 4.2. More specifically, the attackers can guess the private data as the \textit{d} with the maximal \textit{Pr}(\textit{x}$_{\textit{i}}$ = \textit{d}) (for \textit{i} = 1 ${\cdots}$ \textit{M}) by enumeration. Hence, the probability that attackers successfully reverse the \textit{NDB}s and guess the private data is [35]:

\begin{equation}
p_{bf}=\prod_{i=1}^{M}\max_{d=0\dots 2^L-1}Pr\left(x_i=d\right)
\end{equation}

where \textit{d} represents the feasible value of the \textit{i}-th attribute. Since the value of \textit{p}$
_{\textit{bf}}$ is very small, we replace \textit{p}$
_{\textit{bf}}$ with \textit{G} = –log$_{2}$\textit{p}$_{\textit{bf}}$ = –$\sum_{\textit{i} = 1} ^\textit{M} log_2(\textit{max}_{d = 0 {\dots}2^{\textit{L}}-1}\textit{Pr}(\textit{x}_\textit{i} = \textit{d}))$ to analyze the results more clearly. 

We calculate \textit{G} for twelve groups of different \textit{Q}s to observe the security of NegDL for MNIST dataset and CIFAR10 dataset. Similarly, we set \textit{K} = 3, \textit{r} = 6.5, (\textit{p}$_{1}$, \textit{p}$_{2}$, \textit{p}$_{3}$) = (0.70, 0.24, 0.06). Without losing generality, we randomly choose 100 strings for two datasets respectively, and then generate corresponding \textit{NDB}s and calculate \textit{G} and compute the average value for each parameter setting. The results in Table 5 show that the values of \textit{G} for \textit{Q}$_{12}$ is 47 in the case of CIFAR-10, and it indicates that the probability that attackers successfully recover the private data by guessing from an \textit{NDB} is 2$^{-47}$; and \textit{G} = 5082 for \textit{Q}$_{3}$ on CIFAR-10, and it is very low. The \textit{G} for \textit{Q}$_{12}$ is relatively small, but the probability of successfully guessing the private data is still less than 2$^{-40}$. The values of \textit{G} for CIFAR-10 are greater than that for MNIST, because the number of attributes in each channel of instance in CIFAR-10 is 1024 and it is greater than that (i.e. 784) in MNIST. Moreover, it is worth mentioning that attackers have no ways to verify whether they correctly guess the private data according to the sketches, and this further enhances the security of NegDL. We can also observe that the \textit{Q} with better accuracy often has lower \textit{G} value (e.g., the highest level of privacy protection is achieved at \textit{Q}$_{3}$, which corresponds to the lowest accuracy in previous experiments), so the accuracy and security of NegDL are two conflict objectives and we would choose a reasonable balance between security and accuracy according to the requirements in real-world scenarios.

\section{Conclusion }
In this paper, we introduce the negative database to deep learning for privacy protection, and we propose a new model named NegDL. The experimental results on MNIST and CIFAR-10 datasets demonstrate that NegDL can achieve different levels of privacy protection by changing the parameters \textit{Q}. Moreover, NegDL could be comparable to differential privacy in [12]. Compared to original deep learning model without privacy protection, NegDL can preserve most of the accuracy while providing privacy protection.

In future work, we attempt to combine the proposed method with the stochastic gradient descent algorithm and try to protect the weights in distributed deep learning model as well as original private data. We will also try to apply the proposed method to real-world applications involving big data and sensitive information.

\section*{Acknowledgement}
This work is partially supported by the National Natural Science Foundation of China (Grant No. 61806151), the Natural Science Foundation of Chongqing City (Grant No. CSTC2021JCYJ-MSXM4258), and the Key Research and Development Program of Hubei Province (Grant No. 2020AAA001).

\bibliographystyle{unsrt}  


\begin{thebibliography}{1}

\bibitem{1} C. Dwork, “Differential privacy: A survey of results,” in Proceedings of International Conference on Theory and Applications of Models of Computation. Springer, 2008, pp. 1-19.

\bibitem{2} R. L. Rivest, L. Adleman, and M. L. Dertouzos, “On data banks and privacy homomorphisms,” Foundations of Secure Computation, vol. 4, no. 11, pp. 169–180, 1978.

\bibitem{3} F. Esponda, S. Forrest, and P. Helman, “Enhancing privacy through negative representations of data,” Technical Report, New Mexico University, Department of Computer Science, 2004.

\bibitem{4} D. Zhao, X. Hu, S. Xiong, J. Tian, J. Xiang, J. Zhou, and H. Li, “A fine-grained privacy-preserving k-means clustering algorithm upon negative databases,” IEEE Symposium Series on Computational Intelligence. IEEE, 2019, pp. 1945-1951.

\bibitem{5} Y. LeCun, L. Bottou, Y. Bengio, and P. Haffner, “Gradient-based learning applied to document recognition,” Proceedings of the IEEE, vol. 86, no. 11, pp. 2278–2324, 1998.

\bibitem{6} A. Krizhevsky, and G. Hinton, “Learning multiple layers of features from tiny images,” 2009. [Online]. Available: https://citeseerx.ist.psu.edu/viewdoc/download?doi=10.1.1.222.9220\&rep=rep1\&type=pdf

\bibitem{7} Y. Aono, T. Hayashi, L. Wang, and S. Moriai, “Privacy-preserving deep learning via additively homomorphic encryption,” IEEE Transactions on Information Forensics and Security, vol. 13, no. 5, pp.  1333-1345, 2017.

\bibitem{8} Q. Zhu, and X. Lv, “2p-dnn: Privacy-preserving deep neural networks based on homomorphic cryptosystem,” arXiv:1807.08459, 2018.

\bibitem{9} S. Meftah, B. H. M. Tan, C. F. Mun, K. M. Aung, B. Veeravalli, and V. Chandrasekhar, “DOReN: Toward efficient deep convolutional neural networks with fully homomorphic encryption,” IEEE Transactions on Information Forensics and Security, vol. 16, pp. 3740-3752, 2021.

\bibitem{10} J. Park, and H. Lim, “Privacy-preserving federated learning using homomorphic encryption,” Applied Sciences, vol 12, no, 2, pp.734, 2022.

\bibitem{11} R. Shokri, and V. Shmatikov, “Privacy-preserving deep learning,” in Proceedings of the 22nd ACM SIGSAC Conference on Computer and Communications Security. ACM, 2015, pp. 1310-1321.

\bibitem{12} M. Abadi, A. Chu, I. Goodfellow, H. B. McMahan, I. Mironov, K. Talwar, and L. Zhang, “Deep learning with differential privacy,” in Proceedings of the 23nd ACM SIGSAC Conference on Computer and Communications Security. ACM, 2016, pp. 308-318.

\bibitem{13} L. Fan, “Image pixelization with differential privacy”. IFIP Annual Conference on Data and Applications Security and Privacy. Springer, 2018, pp. 148-162.

\bibitem{14} N. Phan, X. Wu, H. Hu, and D Dou, “Adaptive laplace mechanism: Differential privacy preservation in deep learning,” IEEE International Conference on Data Mining. IEEE, 2017, pp. 385-394.

\bibitem{15} C. H. Yu, C. N. Chou, and E. Chang, “Distributed layer-partitioned training for privacy-preserved deep learning,” IEEE Conference on Multimedia Information Processing and Retrieval. IEEE, 2019, pp. 343-346.

\bibitem{16} L. Lyu, J. Yu, K. Nandakumar, Y. Li, X. Ma, J. Jin, and K. S. Ng, “Towards fair and privacy-preserving federated deep models,” IEEE Transactions on Parallel and Distributed Systems, vol. 31, no. 11, pp. 2524-2541, 2020.

\bibitem{17} D. Zhao, and W. Luo, “One-time password authentication scheme based on the negative database,” Engineering Applications of Artificial Intelligence, vol. 62, pp. 396-404, 2017.

\bibitem{18} J. Bringer, and H. Chabanne, “Negative databases for biometric data,” in Proceedings of the 12th ACM Workshop on Multimedia and Security. ACM, 2010, pp. 55-62.

\bibitem{19} D. Zhao, W. Luo, R. Liu, and L. Yue, “Negative iris recognition,” IEEE Transactions on Dependable and Secure Computing, vol. 15, no. 1, pp. 112-125, 2015.

\bibitem{20} D. Zhao, X. Zhou, J. Xiang, and W. Luo, “NDBIris with better unlinkability,” IEEE Symposium Series on Computational Intelligence. IEEE, 2020, pp. 2948-2956.

\bibitem{21} R. Liu, W. Luo, and L. Yue, “Classifying and clustering in negative databases,” Frontiers of Computer Science, vol. 7, no. 6, pp. 864-874, 2013.

\bibitem{22} X. Hu, L. Lu, D. Zhao, J. Xiang, X. Liu, H. Zhou, and J. Tian, “Privacy-preserving k-means clustering upon negative databases,” in Proceedings of International Conference on Neural Information Processing. Springer, Cham, 2018, pp. 191-204.

\bibitem{23} H. Zhu, X. Wang, J. Zhao, S. Geng, and L. Wang, “Multi-client authenticated intersection protocol from a negative database Server,” Journal of Internet Technology, vol. 21, no. 6, pp. 1659-1669, 2020.

\bibitem{24} K. He, X. Zhang, S. Ren, and J. Sun,  “Deep residual learning for image recognition,” in Proceedings of the IEEE Conference on Computer Vision and Pattern Recognition. IEEE, 2016, pp. 770-778.

\bibitem{25} Z.Wang, A. C. Bovik, H. R. Sheikh, and E. P. Simoncelli, “Image quality assessment: from error visibility to structural similarity,” IEEE Transactions on Image Processing, vol. 13, no. 4, pp. 600-612, 2004.

\bibitem{26} A. Hore, and D. Ziou, “Image quality metrics: PSNR vs. SSIM,” International Conference on Pattern Recognition. IEEE, 2010, pp. 2366-2369.

\bibitem{27} F. Esponda, E. S. Ackley, P. Helman, H. Jia, and S. Forrest, “Protecting data privacy through hard-to-reverse negative databases,” International Journal of Information Security, vol. 6, no. 6, pp. 403-415, 2007.

\bibitem{28} D. Zhao, X. Hu, S. Xiong, J. Tian, J. Xiang, J. Zhou, and H. Li, “K-means clustering and kNN classification based on negative databases,” Applied Soft Computing, vol. 110, pp. 1-15, 2021. 


\end{thebibliography}

\end{document}